\documentclass[a4paper]{jpconf}
\usepackage{graphicx}
\usepackage{rotating}
\usepackage{amssymb}
\usepackage{mathptmx}
\usepackage{epsfig}
\usepackage{float}
\usepackage{amssymb,amsmath}
\newcommand{\be}{\begin{equation}}
\newcommand{\ee}{\end{equation}}
\newcommand{\bea}{\begin{eqnarray}}
\newcommand{\eea}{\end{eqnarray}}
\begin{document}

\title{Recoilless Resonant Emission and Detection of Electron Antineutrinos}

\author{W~Potzel}
\address{Physik-Department E15, Technische Universit\"at M\"unchen, James-Franck-Str., 85748 Garching, Germany}

\ead{wpotzel@ph.tum.de}

\begin{abstract}
Recoilless resonant capture of monoenergetic electron antineutrinos (M\"ossbauer antineutrinos) emitted in bound-state $\beta$-decay in the system $^{3}$H - $^{3}$He is discussed. The recoilfree fraction including a possible phonon excitation due to local lattice expansion and contraction at the time of the nuclear transition, homogeneous and inhomogeneous line broadening, and the relativistic second-order Doppler effect are considered. It is demonstrated that homogeneous line broadening is essential due to stochastic magnetic relaxation processes in a metallic lattice. Inhomogeneous line broadening plays an equally important role. An essential issue which has been overlooked up to now, is an energy shift of the resonance line due to the direct influence of the binding energies of the $^{3}$H and $^{3}$He atoms in the lattice on the energy of the electron antineutrinos. This energy shift as well as the second-order Doppler shift exhibit variations in a non-perfect (inhomogeneous) lattice and may seriously jeopardize the observation of M\"ossbauer antineutrinos. If successful in spite of these enormous difficulties, M\"ossbauer antineutrino experiments could be used to gain new and deep insights into the nature of neutrino oscillations, determine the neutrino mass hierarchy as well as up to now unknown oscillation parameters, search for sterile neutrinos, and measure the gravitational redshift of electron antineutrinos in the field of the Earth.
\end{abstract}

\section{Introduction}
After the discovery of the M\"ossbauer effect with $\gamma$-transitions in nuclei \cite{Moessbauer}, the possibility of observing recoilless resonant emission and detection of antineutrinos (M\"ossbauer antineutrinos) was discussed in several publications \cite{Visscher},\cite{Kells}. In Ref. \cite{Kells} an extensive list (including a figure of merit) of possibly suitable nuclear transitions has been provided. More recently, the question concerning M\"ossbauer neutrinos has been revived for the tritium ($^{3}$H) - helium ($^{3}$He) system \cite{RajuRag},\cite{WalterPotzel}. The main idea is to use 18.6 keV electron antineutrinos that are emitted without recoil in the bound-state $\beta$-decay of $^{3}$H to $^{3}$He and are resonantly absorbed - again without recoil - in the reverse bound-state process in which $^{3}$He is transformed to $^{3}$H. Recoilless emission and absorption are intended to be achieved by embedding $^{3}$H and $^{3}$He into metallic lattices.

In the present paper, we want to discuss the basic conditions which have to be fulfilled to enable the observation of the M\"ossbauer effect with antineutrinos in the $^{3}$H - $^{3}$He system. In particular, we will consider lattice expansion and contraction processes which have been neglected up to now but might considerably reduce the recoilfree fraction. We will critically review magnetic relaxation phenomena in metallic systems and show that the induced homogeneous line broadening, as mentioned earlier \cite{WalterPotzel}, can not be avoided, in contradiction to a very optimistic recent suggestion \cite{Raghavan1}. Concerning inhomogeneous broadening we will describe the direct influence of the binding energies of the $^{3}$H and $^{3}$He atoms in the metal matrix on the energy of the antineutrinos - a problem which is discussed here for the first time and which may prevent the observation of M\"ossbauer antineutrinos. We will describe relativistic effects and their connection to second-order Doppler shifts, which again can not be discarded since they also give rise to inhomogeneous line broadening. Both homogeneous and inhomogeneous line broadening effects are estimated to be many orders of magnitude larger than the natural width (minimal width) $\Gamma=\hbar/\tau=1.17\times10^{-24}$ eV as calculated on the basis of the lifetime $\tau=17.81$ y of $^{3}$H. The observation of M\"ossbauer neutrinos will be a very difficult experiment and may turn out to be unsuccessful with the techniques and ideas available at present.

\section{Bound-state $\beta$-decay and its resonant character}

The usual continuum-state $\beta$-decay (C$\beta$) where a neutron in a nucleus transforms into a proton is a three-body process where the emitted electron (e$^-$) and the electron antineutrino ($\bar{\nu_{e}}$) occupy states in the continuum leading to broad energy spectra for the e$^-$ as well as for the $\bar{\nu_{e}}$. In the bound-state $\beta$-decay (B$\beta$), however, the e$^-$ is directly emitted into a bound-state atomic orbit \cite{Bahcall}. Since this is a two-body process, the emitted $\bar{\nu_{e}}$ has a fixed energy $E_{\bar{\nu_{e}}}=Q+B_{z}-E_{R}$. The $\bar{\nu_{e}}$-energy is determined by the Q value, the binding energy $B_{z}$ of the atomic orbit the electron is emitted into, and by the recoil energy $E_{R}$ of the atom formed after the decay. This process occurs, e.g., in the bound-state tritium decay.

The reverse process is also important: a $\bar{\nu_{e}}$ and an e$^-$ in an atomic orbit are absorbed by the nucleus and a proton is transformed into a neutron. Also this is a two-body process. The required energy of the antineutrino is given by $E'_{\bar{\nu_{e}}}=Q+B_{z}+E'_{R}$, where $E'_{R}$ is the recoil energy of the atom after the transformation of a proton into a neutron. This process occurs, e.g., in the bound-state EC-decay of $^{3}$He when irradiated by electron antineutrinos.

Since both $E_{\bar{\nu_{e}}}$ and $E'_{\bar{\nu_{e}}}$ are well defined, B$\beta$ has a resonant character which, however, is partially destroyed by the recoil occurring after emission and absorption of the $\bar{\nu_{e}}$. The resonance cross section is given by \cite{Mika} $\sigma=4.18\cdot10^{-41}\cdot g_{0}^{2}\cdot \varrho(E_{\bar{\nu_{e}}}^{res})/ft_{1/2}$ (in units of cm$^{2}$) where $g_{0}=4\pi(\hbar/mc)^{3}\vert\psi\vert^{2}\approx4(Z/137)^{3}$ for low-Z, hydrogen-like wavefunctions $\psi$; \textit{m} is the electron mass, and $\varrho(E_{\bar{\nu_{e}}}^{res})$ is the resonant spectral density, i.e., the number of antineutrinos in an energy interval of 1 MeV around $E_{\bar{\nu_{e}}}^{res}$. For a super-allowed transition, $ft_{1/2}\approx1000$ s. For the $^{3}$H - $^{3}$He system, the resonance energy is 18.60 keV, $ft_{1/2}=1000$ s, and B$\beta/$C$\beta=6.9\cdot10^{-3}$ with 80\% and 20\% of the B$\beta$ events proceeding via the atomic ground state and excited atomic states, respectively \cite{Bahcall}. If gases of $^{3}$H and $^{3}$He are used at room temperature the profiles of the emission and absorption probabilities are Doppler broadened and both emission as well as absorption of the electron antineutrinos will occur with recoil. Thus the expected resonance cross-section $\sigma\approx 1\cdot10^{-42}$ cm$^{2}$. Clearly, an observation would require very strong sources of $^{3}$H and large target ($^{3}$He) masses \cite{RajuRag} and thus make such an experiment virtually impossible. However, making use of the M\"ossbauer effect of antineutrinos, i.e., using recoilfree resonant antineutrino emission and absorption, would increase the resonance cross-section by many orders of magnitude, typically to $\sigma_{R}\approx 1\cdot10^{-33}$ cm$^{2}$ \cite{RajuRag},\cite {WalterPotzel}.

\section{M\"ossbauer antineutrinos: Recoilfree resonant antineutrino emission and absorption} 

\subsection{Recoilfree fraction}
\subsubsection{Lattice excitations due to momentum transfer}
In analogy to usual M\"ossbauer spectroscopy with photons \cite{Moessbauer},\cite{Frau}, recoilfree emission and absorption can be realized by embedding the atoms of the source (in the present example, $^{3}$H) and the target ($^{3}$He) into solid-state lattices, e.g., into metallic matrices. Recoilfree processes require that the lattice excitations (e.g. phonons) will remain unchanged by the emission and the absorption of the antineutrino. Considering the momentum transfer only, the recoilfree fraction is given by $f=exp \left\{-(\frac{E}{\hbar c})^{2}\cdot\left\langle x^{2}\right\rangle\right\}$, where $E$ is the transition energy (18.6 keV for the $^{3}$H - $^{3}$He system) and $\left\langle x^{2}\right\rangle$ is the mean-square atomic displacement. The recoilfree fraction is biggest at low temperatures. However, even at very low temperatures, $f<1$, because of the zero-point motion, which itself is a consequence of the Heisenberg uncertainty principle. In the Debye approximation and in the limit of very low temperatures T, $f(T\rightarrow0)=exp \left\{-\frac{E^{2}}{2Mc^{2}}\cdot\frac{3}{2k_{B}\theta}\right\}$, where $\theta$ is the effective Debye temperature, $k_{B}$ is the Boltzmann constant, and $\frac{E^{2}}{2Mc^{2}}$ is the recoil energy which would be transmitted to a free atom of mass $M$. For $^{3}$H and $^{3}$He in a metal matrix, e.g. Nb, effective Debye temperatures up to $\theta\approx800$ K have been estimated \cite{RajuRag},\cite{Raghavan1}. Thus $f(0)\approx0.27$ and the probability for recoilfree emission and consecutive recoilfree capture of electron antineutrinos is given by $f^{2}\approx0.07$ (for $T\rightarrow0$).

\subsubsection{Lattice excitations by lattice expansion and contraction}
Concerning the recoilfree fraction, the anology with usual (photon) M\"oss\-bauer spectroscopy is insufficient: For electron antineutrinos one has to consider an additional process, which is not important for photon M\"ossbauer spectroscopy with excited nuclear states. In this latter case, a radioactive parent nucleus decays and populates a nuclear level (the M\"ossbauer state) of the daughter nucleus. Deexcitation from this M\"ossbauer state to the groundstate can then proceed via the recoilfree emission of a M\"ossbauer photon. Due to the relatively long lifetime of the M\"ossbauer state, most rearrangement processes of the electron shell caused by the preceeding nuclear decay and occurring on a timescale of $<10^{-12}$ s (much shorter than the lifetime of the M\"ossbauer state) have settled.\footnote{An exception are so-called after effects in non-metallic lattices where some rearrangement processes in the electron shell involve rather long-lived (several ns) electronic configurations, e.g., different oxydation states, which can be detected in M\"ossbauer spectroscopy by their different hyperfine interactions \cite{Wickman}.} Thus, the emitted M\"ossbauer photon from the source is not directly involved in the nuclear decay and, therefore, is likely to exhibit the right energy that is required to excite the same M\"ossbauer level in the absorber (target). The situation is different for antineutrino emission and capture. These processes occur at the same time when the nuclear transformation to the different element takes place; in fact the antineutrinos take part in the {\em nuclear} transformation processes. Considering the $^{3}$H - $^{3}$He system as the particular example, $^{3}$H is more strongly bound in the metallic lattice than $^{3}$He, and $^{3}$H and $^{3}$He use different amounts of space. As a consequence, the lattice-deformation energies for $^{3}$H and $^{3}$He, e.g. in the Nb lattice, are $E_L^{^{3}H}=0.099$ eV and $E_L^{^{3}He}=0.551$ eV, respectively \cite{Puska}. Assuming again an effective Debye temperature of $\theta\approx800$ K one can estimate - in analogy to the situation with momentum transfer - that the probability that this lattice deformation will {\em not} cause lattice excitations is smaller than $exp \left\{-\frac{E_L^{^{3}He}-E_L^{^{3}H}}{k_{B}\theta}\right\}\approx1\cdot10^{-3}$. This factor appears in the emission as well as in the capture process of the antineutrinos. Thus, in addition to the factor $f^{2}\approx0.07$ (for $T\rightarrow0$), which reflects the momentum transfer in the emission and capture processes of the $\bar{\nu_{e}}$, lattice deformation results in a further reduction factor of at least $1\cdot10^{-6}$.

Therefore the total probability for recoilfree emission and consecutive recoilfree capture of $\bar{\nu_{e}}$ is smaller than $\approx7\cdot10^{-8}$. This estimate has to be confirmed by more detailed theoretical calculations; however, it emphasizes the importance of a thorough understanding of $\bar{\nu_{e}}$ recoilfree emission and capture.

\subsection{Linewidth}

For the $^{3}$H decay with a lifetime $\tau=17.81$ y, the natural linewidth $\Gamma$, i.e., the minimal width, $\Gamma=\hbar/\tau=1.17\times10^{-24}$ eV. In two recent contributions \cite{Raghavan1} it has been argued that such a narrow line can indeed be achieved experimentally. We strongly disagree with these claims and want to stress that basically two types of line broadening are important \cite{Balko},\cite{Coussement},\cite{WalterPotzeletal},\cite{WalterPotzel}.

\subsubsection{Homogeneous broadening}
Homogeneous broadening is caused by electromagnetic relaxation. For example, spin-spin interactions between nuclear spins of $^{3}$H and $^{3}$He and with the spins of the nuclei of the metallic lattice lead to fluctuating magnetic fields. Contrary to the notion in \cite{Raghavan1}, magnetic relaxations are stochastic processes and can not be described by a periodic energy modulation of an excited hyperfine state. The simplest magnetic relaxation model consists of a three-level system: the groundstate and two excited hyperfine-split states (energy separation $\hbar\Omega_{0}$) between which transitions (so-called relaxation processes) take place with an average frequency $\Omega$. With stochastic processes, three frequency regimes can be distinguished \cite{WickmanHH}:

a) $\Omega\ll\Omega_{0}$. For the simple three-level system, two lines separated by $\hbar\Omega_{0}$ will be observed. The lines will be broadened to an effective experimental linewidth $\Gamma_{exp}\approx\hbar\Omega$ as suggested by the time-energy uncertainty principle. Only in the limit of very small $\Omega$ values, the lines exhibit natural width. With increasing $\Omega$ the lines broaden.

b) $\Omega\approx\Omega_{0}$. The lines are severely broadened. In fact, the intensity is distributed over a broad pattern which extends roughly over a range given by the total hyperfine splitting $\hbar\Omega_{0}$, as suggested by the time-energy uncertainty principle.

c) $\Omega\gg\Omega_{0}$. This is the frequency regime of motional narrowing. The system stays only for a short time (typically $1/\Omega$) in one of the levels and then stochastically jumps into the other one. Thus an averaging process over the energies of both levels takes place, resulting in one line at the center of the hyperfine-splitting pattern. In this high-frequency limit the linewidth is practically natural.

For typical hyperfine splittings due to nuclear spin-spin interaction in metallic lattices, $\Omega_{0}\approx10^{5}$ s$^{-1}$. Typical relaxation times for $^{3}$H and $^{3}$He in a Pd lattice are $T_{2}\approx2$ ms, and for NbH, $T_{2}\approx79$ $\mu$s \cite{Stoll},\cite{RajuRag}. The latter gives a linewidth $\Gamma_{exp}$ due to homogeneous broadening, $\Gamma_{exp}=9\times10^{-12}$ eV $\approx7\times10^{12}\Gamma$. Due to stochastic relaxation processes, homogeneous broadening by $\approx12$ orders of magnitude has to be expected in the $^{3}$H - $^{3}$He system because the stochastic relaxation frequences are far below the motional-narrowing regime. As a consequence, also the resonance cross-section $\sigma_{R}$ will be reduced by $\approx12$ orders of magnitude. To compensate for this effect one would have to increase the number of $^{3}$He atoms in the target by the same factor!

\subsubsection{Inhomogeneous broadening}
In imperfect lattices, inhomogeneous broadening is caused by stationary effects, in particular by impurities, lattice defects, variations in the lattice constant, and other effects which destroy the periodicity of the lattice. This is a critical issue also for $\bar{\nu_{e}}$ since - as will be described in section 4 -  the source contains a large amount of $^{3}$H but very little $^{3}$He whereas the target contains a lot of $^{3}$He
but practically no $^{3}$H \cite{RajuRag},\cite{WalterPotzel}. Clearly, the $^{3}$H and $^{3}$He distributions on the interstitial sites will be random and will destroy lattice periodicity. An additional process due to variations of the zero-point energy will be discussed in section 3.3. In general, for photon M\"ossbauer spectroscopy, in the best single crystals, inhomogeneous broadening is of the order of $10^{-13}$ eV to $10^{-12}$ eV \cite{Coussement}.

A feature which is different between usual M\"ossbauer spectroscopy with photons and M\"ossbauer antineutrinos is connected to the variation of the binding energies of $^{3}$H and $^{3}$He in an inhomogeneous metallic lattice. The binding energies directly affect the energy of the $\bar{\nu_{e}}$: In a perfect lattice, the difference in binding energies in the lattice between $^{3}$He and $^{3}$H \cite{Puska}, which is given to the $\bar{\nu_{e}}$ when $^{3}$H decays into $^{3}$He, exactly compensates the binding-energy difference needed for the reverse process, when the $\bar{\nu_{e}}$ is captured and $^{3}$He transforms into $^{3}$H. However, in the source and target intended to be used in a real experiment (see section 4) such a compensation even within an experimental linewidth of $\Gamma_{exp}=9\times10^{-12}$ eV will be extremely unlikely considering the fact that the binding energies per atom are in the eV range \cite{Puska}. Inhomogeneities in real lattices as mentioned above and, in particular, the vastly different amounts of $^{3}$H and $^{3}$He in the source and in the target, will result in variations of the binding energies of the $^{3}$H and $^{3}$He atoms in source and target much larger than $\Gamma_{exp}$ and thus destroy the resonance condition.

In usual M\"ossbauer spectroscopy with photons, different binding strengths due to inhomogeneities in source and absorber (target) affect the photon energy only via the {\em change} in the mean-square nuclear charge radius between the groundstate and the excited state of the nucleus. This leads to the {\em isomer shift}, i.e., a shift of the photon energy in the range typical for {\em hyperfine interactions} (in the $10^{-7} - 10^{-9}$ eV range) \cite{ShenoyWagner}. Thus variations of isomer shifts due to an inhomogeneous lattice lead to a line broadening which is also in the neV range. For $\bar{\nu_{e}}$ of the $^{3}$H - $^{3}$He system, a neV shift would already correspond to $\sim100\Gamma_{exp}\approx10^{15}\Gamma$. However, since the $\bar{\nu_{e}}$ energy is {\em directly} affected, one has to expect a variation of the $\bar{\nu_{e}}$ energy several orders of magnitude larger than the neV range, i.e., much larger than $10^{15}\Gamma$.

\subsection{Relativistic effects}

An atom vibrating around its equilibrium position in a lattice does not only exhibit a mean-square displacement $\left\langle x^{2}\right\rangle$ but also a mean-square velocity $\left\langle v^{2}\right\rangle$. According to Special Relativity Theory this causes a time-dilatation effect which results in a reduction of frequency (energy): $\Delta\omega=\omega-\omega'=-v^{2}\omega/(2c^{2})$. Since this reduction is proportional to $(v/c)^{2}$ it is often called second-order Doppler shift (SOD).

Within the Debye model, the energy shift between source (s) at temperature $T_{s}$ and target (t) at temperature $T_{t}$ is given by

\bigskip
$(\Delta E/E)=\frac{9k_{B}}{16Mc^{2}}(\theta_{s}-\theta_{t})
+\frac{3k_{B}}{2Mc^{2}}\left[T_{s}\cdot f(T_{s}/\theta_{s})-T_{t}\cdot f(T_{t}/\theta_{t})\right]$

\smallskip
where $f(T/\theta)=3(\frac{T}{\theta})^{3}\cdot\int^{\theta/T}_{0}\frac{x^{3}}{exp(x)-1}dx$.

\bigskip
At low temperatures, the temperature-dependent term can be neglected if source and target are at about the same temperature (e.g., in a liquid-He bath at 4.2 K) \cite{WalterPotzel}. However, even in the low-temperature limit, the first term which is caused by the zero-point energy can not be neglected. As discussed in detail in \cite{WalterPotzel}, it is {\em not} required that the chemical bonds (i.e. the Debye temperatures) of $^{3}$H and $^{3}$He have to be the same in the metal matrix. However, the chemical bond for $^{3}$H has to be the same in source and target, and the same condition has to be fulfilled for $^{3}$He. As a consequence, the surroundings of $^{3}$H (and those for $^{3}$He) including nearest and higher-order nearest neighbors in source and target should be as similar as possible. This is critical, since, as already mentioned in section 3.2, source and target contain vastly different amounts of $^{3}$H and $^{3}$He \cite{RajuRag},\cite{WalterPotzel}.

Another critical issue is a possible variation of the effective Debye temperature due to inhomogeneities within the lattice of the source (or the target) itself. If the effective Debye temperature varies by 1 K, the relative change due to a variation of the zero-point energy would be $(\Delta E/E)\approx2\times10^{-14}$ which corresponds to a lineshift of $3\times10^{14}$ times the natural width $\Gamma$. Even after taking homogeneous broadening into account, i.e., considering the above-mentioned linewidth $\Gamma_{exp}=9\times10^{-12}$ eV, the variation of the zero-point energy would still be $\sim43\Gamma_{exp}$. Thus, also the variation of the zero-point energy within the source and the target will significantly contribute to inhomogeneous broadening.

\section{Feasibility of M\"ossbauer antineutrino ex\-periments}
\subsection{The $^{3}$H - $^{3}$He system}

As discussed in detail in \cite{Kells},\cite{RajuRag},\cite{WalterPotzel}, the most promising system for the observation of M\"ossbauer antineutrinos is $^{3}$H as source and $^{3}$He as target. The key idea is to use Nb as metal matrix \cite{RajuRag}. The source can be produced by chemically loading $^{3}$H into metallic Nb, where $^{3}$H occupies the tetrahedral interstitial sites. The $^{3}$He target can be prepared by using the 'tritium trick': First load $^{3}$H into the Nb matrix, wait until enough $^{3}$H has decayed into $^{3}$He, and then remove all the remaining $^{3}$H from the Nb matrix. After this procedure, also the $^{3}$He should occupy the tetrahedral interstitial sites.

For a base line of $10$ m, typical source strengths of 1 MCi of $^{3}$H and absorber masses of $\sim1$ g of $^{3}$He have been considered resulting in a typical $\beta$-activity of $\sim10$ counts per day after an activation time of 65 days \cite{RajuRag}.

\subsection{Enormous experimental difficulties}
The most difficult experimental problems to observe recoilfree resonant emission and absorption of $\bar{\nu_{e}}$ are the following:
\begin{enumerate}
\item \textit{How large is the recoilfree fraction?} The recoilfree fraction due to the momentum transfer of the $\bar{\nu_{e}}$ is reasonable: $f^{2}\approx0.07$ (for $T\rightarrow0$). However, the critical issue is the lattice expansion and contraction of the metallic lattice due to the different elements before and after the decay (see section 3.1.2): An additional reduction factor of at least six orders of magnitude has been estimated. This problem has been overlooked until now, because it does not play a role in the M\"ossbauer effect with photons.
\item \textit{How small an experimental linewidth can be achieved?} In contrast to the claims presented in \cite{Raghavan1}, the natural linewidth $\Gamma=1.17\times10^{-24}$ eV will not be observed, mainly because of two reasons: a) {\em Homogeneous line broadening} will be caused by {\em stochastic} magnetic relaxation. The effect of other processes connected with lattice vibrations are unimportant.\footnote{In the frequency modulation picture, lattice vibrations cause sidebands, the amplitudes of which are tiny, because the modulation index is small due to the high frequencies of the lattice vibrations. Thus, essentially only the carrier frequency remains. In cases where the modulation index is big and sidebands with appreciable amplitudes are expected, the recoilfree fraction vanishes, since the mean-square atomic displacement is large, i.e., there is no M\"ossbauer effect \cite{Frau}.} Homogeneous broadening is estimated to cause an experimental linewidth $\Gamma_{exp}=9\times10^{-12}$ eV $\approx7\times10^{12}\Gamma$ \cite{WalterPotzel}. b) {\em Inhomogeneous broadening} can not be avoided in such imperfect lattice systems as $^{3}$H in Nb and $^{3}$He in Nb: The random distributions of $^{3}$H and $^{3}$He in source and target destroy the lattice periodicity, and different contents of these elements will cause further problems. Thus, in addition to second-order Doppler shifts (see section 3.3) of the resonance lines due to (small) changes in the zero-point energy, much larger shifts are expected because of the variation of binding energies (strengths) of $^{3}$H and $^{3}$He in the imperfect Nb metal matrix (see section 3.2.2). These different binding energies {\em directly} influence the energy of the $\bar{\nu_{e}}$. Also this shift has been overlooked until now, because it is only a tiny (hyperfine interaction) effect leading to the isomer shift in M\"ossbauer spectroscopy with photons. With M\"ossbauer antineutrinos, this {\em direct} influence on the $\bar{\nu_{e}}$ energy will cause a broadening much larger than $10^{15}\Gamma$ (and thus $\sigma_{R}$ will be reduced by the same factor, see section 3.2) and can turn out to be the main killer of such an experiment.

\item \textit{Can $^{3}$H be removed from the Nb metal matrix to low enough concentrations?} The 'tritium trick' is a very clever method to make sure that both $^{3}$H and $^{3}$He occupy the tetrahedral interstitial sites in the Nb metal matrix \cite{RajuRag}. However, any $^{3}$H remaining in the target will act as unwanted background.

\item \textit{The application of low temperatures (liquid He)} will be necessary to achieve temperature stability and to make sure that source and target are at the same temperature. Otherwise second-order Doppler shifts due to different temperatures will be present. Another requirement is an efficient and homogeneous (across the source matrix) dissipation of the heat generated by non-resonant C$\beta$ decay events in the source.

\end{enumerate}

\section{Interesting experiments}
If M\"ossbauer antineutrinos could be observed successfully, several interesting experiments could be performed. Here is a short selection:

\begin{enumerate}
\item A M\"ossbauer antineutrino experiment could provide a unique possibility for a better understanding of the true nature of neutrino oscillations \cite{Bilenky},\cite{Bilenkyetal1},\cite{Bilenkyetal2},\cite{Akhmedov}. Because of the extremely sharp energy distribution, oscillations of M\"ossbauer antineutrinos would indicate a stationary phenomenon where the evolution of the neutrino state occurs in space rather than in time.
\item Due to the low $\bar{\nu_{e}}$ energy of 18.6 keV, M\"ossbauer-antineutrino oscillations would allow us to use ultra-short base lines to determine oscillation parameters and the mass hierarchy. For example, for the determination of the mixing angle $\Theta_{13}$, a base line of only $\sim10$ m (instead of 1500 m) would be sufficient. In addition, very small uncertainties for $\Theta_{13}$, accurate measurements of $\Delta m^{2}_{12}$ and $\Delta m^{2}_{31}$ \cite{Minakata}, and a determination of the neutrino mass hierarchy \cite{Nunokawa} could be achieved.
\item In a disappearance experiment with 18.6 keV M\"ossbauer $\bar{\nu_{e}}$ a search for the conversion $\bar{\nu_{e}}$ $\rightarrow\nu_{sterile}$ \cite{Kopeikin} could be performed. If $\Delta m^{2}\approx1$ eV, the oscillation length would only be $\sim5$ cm. This would require ultra-short base lines, which would be difficult to realize otherwise.
\item Gravitational redshift $\bar{\nu_{e}}$ measurements within a distance of $\sim4$ m in the gravitational field of the Earth could be carried out. In contrast to photons, $\bar{\nu_{e}}$ are particles with a (small) rest mass and, in principle, could behave differently from photons in a gravitational field. For the determination of the redshift an accuracy of $\sim0.01\times\Gamma_{exp}$ could be reached.

\end{enumerate}

\section{Conclusions}

The system $^{3}$H (source) and $^{3}$He (target) has been considered for a possible observation of recoilfree resonant (M\"ossbauer) emission and absorption of electron antineutrinos. M\"ossbauer antineutrinos would allow us to perfom interesting new measurements. However, the experiment is very challenging as emphasized in sections 3 and 4. In particular, we have argued that - contrary to the claims of Ref. \cite{Raghavan1} - the natural linewidth ($\Gamma=1.17\times10^{-24}$ eV) will not be observed because homogeneous broadening alone would result in an experimental linewidth of $\Gamma_{exp}=9\times10^{-12}$ eV $\approx7\times10^{12}\Gamma$. Even if an experimental linewidth of $\Gamma_{exp}=9\times10^{-12}$ eV could be reached (which would still be highly fascinating for many $\bar{\nu_{e}}$ experiments), the chances for the observation of M\"ossbauer antineutrinos are drastically reduced because of two newly discussed effects. They have been overlooked up to now because they play no significant role with photon M\"ossbauer spectroscopy: a) an additional suppression of the recoilfree fraction because of {\em lattice expansion and contraction at the time of the nuclear transition}, and b) the {\em direct influence of the binding energies} of $^{3}$H and $^{3}$He atoms in the metal matrix on the energy of the electron antineutrino. The variation of these binding energies in the inhomogeneous metal matrix can lead to lineshifts and thus to line broadenings much larger than $10^{15}\Gamma$. Theoretical lattice-dynamics calculations are required to better understand and estimate the importance of both effects. Still, it is obvious already now that both effects might prevent the observation of M\"ossbauer antineutrinos.

\ack
It is a pleasure to thank F. Wagner and S. Roth, Physik-Department E15, Technische Universit\"at M\"unchen, for numerous and very fruitful discussions. This work has been supported by funds of the Deutsche Forschungsgemeinschaft DFG (Transregio 27: Neutrinos and Beyond), the Munich Cluster of Excellence (Origin and Structure of the Universe), and the Maier-Leibnitz-Laboratorium (Garching).

\end{document}